\documentclass[screen,acmsmall,table]{acmart}

\usepackage{libertine}
\usepackage{xcolor}
\usepackage{balance}
\usepackage{color}
\usepackage{booktabs} 
\usepackage{url}
\usepackage{graphicx}
\usepackage{subfigure}
\usepackage{multirow}
\usepackage{xspace}
\usepackage{enumitem} 
\usepackage[normalem]{ulem}
\usepackage{soul}
\usepackage{balance}
\usepackage{comment}
\usepackage{amsmath}
\usepackage{microtype}
\usepackage{geometry}
\usepackage{tcolorbox}
\usepackage[frozencache,cachedir=minted-cache]{minted}

\newcommand{\han}[1]{\textcolor{black}{#1}}

\newcommand{\tool}{{\texttt{CLAP}}\xspace}

\setcopyright{none}
\author{Han Wang}
\email{han.wang@monash.edu}
\affiliation{%
  \institution{Faculty   of   Information   Technology,   Monash   University}
  \country{Australia}
}
\author{Han Hu}
\email{han.hu@monash.edu}
\affiliation{%
   \institution{Faculty   of   Information   Technology,   Monash   University}
  \country{Australia}
}
\author{Chunyang Chen}
\email{chun-yang.chen@tum.de}
\affiliation{%
  \institution{Technical University of Munich \& Monash University}
  \country{Germany}
}

\author{Burak Turhan}
\email{burak.turhan@oulu.fi}
\affiliation{%
  \institution{M3S Research Unit, Faculty of ITEE, University of Oulu}
  \country{Finland}
}

\begin{document}

\title{Chat-like Asserts Prediction with the Support of Large Language Model}


\begin{abstract}
Unit testing is an essential component of software testing, with the assert statements playing an important role in determining whether the tested function operates as expected. Although research has explored automated test case generation, generating meaningful assert statements remains an ongoing challenge. While several studies have investigated assert statement generation in Java, limited work addresses this task in popular dynamically-typed programming languages like Python. In this paper, we introduce Chat-like execution-based Asserts Prediction (\tool), a novel Large Language Model-based approach for generating meaningful assert statements for Python projects. \tool utilizes the persona, Chain-of-Thought, and one-shot learning techniques in the prompt design, and conducts rounds of communication with LLM and Python interpreter to generate meaningful assert statements. We also present a Python assert statement dataset mined from GitHub. Our evaluation demonstrates that \tool achieves 64.7\% accuracy for single assert statement generation and 62\% for overall assert statement generation, outperforming the existing approaches. We also analyze the mismatched assert statements, which may still share the same functionality and discuss the potential help \tool could offer to the automated Python unit test generation. The findings indicate that \tool has the potential to benefit the SE community through more practical usage scenarios.

\end{abstract}
\keywords{Unit Testing, LLM, Asserts Prediction}

\ccsdesc[500]{Software and its engineering~Software testing and debugging}


\maketitle


\section{Introduction}
\label{sec:introduction}

Unit testing plays an important role in software testing, as it can detect and diagnose bugs in the existing system by testing small units at the early stage~\cite{trautsch2017there,daka2014survey}. A unit test case normally consists of test inputs and test oracles (assert statements)~\cite{yu2022automated}. The assertions are crucial in the unit testing process, as they compare the actual and expected outputs to determine whether a tested function performs as intended.  High-quality assert statements enhance the maintainability of unit tests, help better identify bugs, and assist future developers in understanding the purpose and output of the code. In addition, if a test case fails, the assert statements serve as a valuable reference for identifying and resolving faults.

Developing unit tests for a project can be a demanding and time-consuming task for developers~\cite{lukasczyk2023empirical, whittaker2000software}. In response to this challenge, automated unit test generation has emerged as a well-established field of research. Studies have concentrated on automating unit test generation using rule-based or search-based algorithms for both Java and Python projects~\cite{lukasczyk2022pynguin, fraser2011evosuite, maciver2020test, taneja2008diffgen}. 
However, while search-based testing approaches can facilitate the automation of test input generation and improve test coverage, generating complete and meaningful assert statements that meet the actual needs of developers remains a challenge~\cite{ watson2020learning}. To be specific, these approaches either capture and assert the return values of non-void-return methods or introduce mutants into the software and attempt to generate assert statements able to kill these mutants. 
Interviews~\cite{almasi2017industrial} with developers have revealed that the automatically generated assertions are often poor and simple, resulting in their inability to detect real bugs, and the lack of meaningfulness in these assertions leads developers to prefer handwritten test cases over the generated ones, ultimately undermining the purpose of automated testing~\cite{watson2020learning}.

Therefore, recent works have started to explore automatically generating meaningful assert statements in Java given the input of the test input and its focal method (method to be tested)~\cite{watson2020learning, yu2022automated, nashidretrieval, tufano2022generating,dinella2022toga, mastropaolo2021studying}. 
However, for dynamically typed programming languages such as Python, though it has become the most popular programming language~\cite{web:languageRanking}, no previous research has investigated the generation of meaningful assert statements. There is a demand for more advanced and effective tools to assist developers in this area and enhance the overall code quality. Unlike Java, Python is dynamically typed, features flexible syntax, and employs high-level abstractions, making it more challenging to accurately capture the expected behaviour or values of the focal method. 
\han{Directly adopting current approaches to assert statement generation in Python has limitations. First, these deep learning (DL) or information retrieval (IR) based methods rely on a training set, and performance varies with the set's size and relevance. Unfortunately, Python lacks a standardised dataset specifically for assert generation, complicating the adoption of these models. Python's diverse testing practices and dynamic typing system, is challenging for automated test identification and assert extraction, further compounded by a less mature testing tools ecosystem that limits the availability of training data for such methods.}
Second, we found that 71.8\% of the test cases we crawled from open-source Python projects contain more than one assert statement in a test case, which is not supported by current approaches designed only for a single assert statement. \han{Generating multiple assertions using the current single assert statement can be challenging as identified in previous work~\cite{watson2020learning}. These methods typically treat each assertion independently, but fail to consider the connections/dependencies of the asserts in the same test cases.} These limitations highlight the need for new approaches to generate multiple meaningful assert statements for Python unit testing.

In recent years, Large Language Models (LLM), such as GPT~\cite{web:gpt} and LLAMA~\cite{touvron2023llama, touvron2023llama2}, have emerged as pre-trained neural network models capable of achieving impressive results in various tasks 
such as code completion~\cite{nguyen2022empirical, li2023skcoder, maddigan2023chat2vis, web:Copilot}, documentation and comments generation~\cite{khan2022automatic, geng2023empirical}, repair programs and fix bugs~\cite{nashidretrieval, prenner2022can, cao2023study}. 
Leveraging LLMs' pre-trained nature and their proficiency in understanding code context, we use them to automate the assert generation in Python.
Notably, GPT models perform well at Python code generation compared to other languages, likely due to their pre-release training set distribution~\cite{cassano2023multipl, web:gpt}.

In this paper, we propose \tool (\textbf{C}hat-like \textbf{L}LM-based \textbf{A}sserts \textbf{P}rediction), a novel LLM-based approach for generating meaningful assert statements for Python projects. We adopt state-of-art prompt design techniques, such as one-shot learning and Chain-of-Thought (CoT), into the assert generation prompts. To include more context of the given code, we design three prompt templates to communicate with the LLM as well as support the multi asserts generation given one test case. We execute the initial round of predictions and relay any error messages received from the interpreter back to the LLM, prompting it to correct the mistakes and refine its predictions.
We also mine GitHub and release an open-source Python assert statement dataset. Lastly, we implement the approach with OpenAI LLMs and evaluate it with the dataset. 

We evaluate the effectiveness of our prompt design and compare \tool performance with state-of-the-art approaches~\cite{watson2020learning, nashidretrieval, wang2021codet5}. While the prompt design demonstrates strong performance with the two OpenAI LLMs, it achieves better results on the \textit{text-davinci-003}~\cite{web:gpt}. The prompt techniques we integrate into \tool can help the LLM generate more accurate assertions. In our evaluation, \tool attains 64.7\% accuracy for single assert statement generation,  and achieves 62\% accuracy in generating all assert statements that no other approaches have explored. We collect the generated assertions that can improve the test code quality or fit more to the test purpose and submit pull requests to the tested projects, and 7 out of 9 pull requests has been approved.

In summary, the contribution of this work is:
\begin{itemize}
    \item The study specifically focuses on assert generation in Python, encompassing multi-assert generation and the utilization of various testing libraries. 
    \item \tool, a system that integrates rounds of CoT and feedback prompts to interact with the LLM, effectively addressing the assert generation task.
    \item We make both our tool, \tool, and our Python assert dataset publicly available for the benefit of the research community~\cite{web:clap}.
\end{itemize}
\section{Background}
\label{sec:background}

\subsection{Unit Testing \& Assert Statements}
\label{sec:background1}
A unit test is a small and executable piece of code that exercises the functionality of a unit under test~\cite{almasi2017industrial}. Unit testing plays an important role in test-driven development (TDD) as it can better help with fault detection~\cite{tosun2018effectiveness}, contribute to developer's productivity~\cite{fucci2014role}, and can also serve as documentation and specification~\cite{shamshiri2015automatically}. The existence of unit tests ensures the correctness, reliability, and maintainability of the units within the code~\cite{yoshida2016fsx, gonzalez2017large}. 

\han{To implement unit tests, developers use specific conventions and frameworks that define how test methods are structured and recognized within different programming environments. 
In Java projects, test methods start with $@Test$ annotation (inherent to the JUnit framework)~\cite{watson2020learning}. However, in Python, projects do not have a universal annotation for indicating test methods. Instead, they often rely on naming conventions and specific testing frameworks such as unittest~\cite{web:pythonunittest}. In Python, test methods are generally prefixed with $test$~\cite{trautsch2017there} and contained modules following the pattern $test\_module\_name.py$ or $module\_name\_test.py$.}

Test oracles are important parts of the test methods, normally appearing as assert statements. They validate whether the given conditions are met during the execution. 
In Python, there are two common ways to write assert statements. The first method involves utilising the assert keywords provided by Python, which allows for limited but straightforward checks of conditions within the code. The second approach entails leveraging other testing libraries, such as unittest~\cite{web:pythonunittest}, scikit-learn~\cite{web:sklearn} and tf.test~\cite{web:tfunittest}, which offer over 70 specialized assert statements and additional features for a wide range of testing scenarios. For example, the \textit{assertRaisesRegex} checks if a specific exception is raised during the execution and if the exception message matches the regular expression pattern. Therefore, the assert statements in Python vary depending on the specific requirements and preferences of developers.

\subsection{Automatic Assert Generation}

\han{Automatic assert generation is a critical challenge in software testing. It aims to automate the creation of assert statements within unit tests. The challenge in automatic assert generation lies in effectively predicting assert statements that accurately reflect the intended functionality of the application.}

\han{The pioneering work in this area was introduced by Watson et al.~\cite{watson2020learning} in 2020. They developed a Deep Learning (DL) approach using a Neural Machine Translation (NMT) model, named ATLAS, to generate assert statements specifically for Java. 
They proposed the Test-Assert Pairs(TAPs) dataset extracted from Java unit test method. This dataset paired unit tests with their corresponding focal methods, providing a structured way to train and evaluate the model. Since then, all works have followed this structure, generating assert statements based on the given rest of the unit test and its focal method.
Subsequent developments~\cite{yu2022automated, tufano2022generating, mastropaolo2021studying, mastropaolo2022using, nashidretrieval} have enhanced the assert generation in Java with mixed methods, including information retrieval and the tuning of transformer and large language models. Beyond Java, developments in other languages like JavaScript have also progressed; for instance, Zamprogno et al.~\cite{zamprogno2022dynamic} introduced AutoAssert, which incorporates a human-in-the-loop mechanism. The advancements highlight the increasing interest in automatic assert generation, aimed at reducing manual efforts.
}

\subsection{Large Language Model \& Prompt Design}

Large Language Models (LLMs) like GPT, with over 175 billion parameters, excel in diverse tasks such as summarizing and code completion. LLMs are Trained on extensive datasets, including various open-source projects, and built on the Transformer architecture~\cite{vaswani2017attention}. They utilise mechanisms like self-attention to comprehend prompts and capture contextual relationships between words from different distances.
To control the output of LLMs, prompts are fed into the models. Prompt design influences the LLM's ability to capture the requirements and produce specific and coherent results. The concept of prompt engineering~\cite{liu2022design} has been proposed to describe the iterative process of crafting and refining prompts to optimize the performance of LLMs in generating desired outputs. The process may include incorporating natural language descriptions, such as ``\textit{The aim of this task is to xxx}'', or give the LLMs one or a few samples (one-shot or few-shots learning) to better understand the task and guide the model in formatting the output content.

 \section{Methodology}
\label{sec:methodology}

\begin{figure*}
    \centering
	\includegraphics[width=1\textwidth]{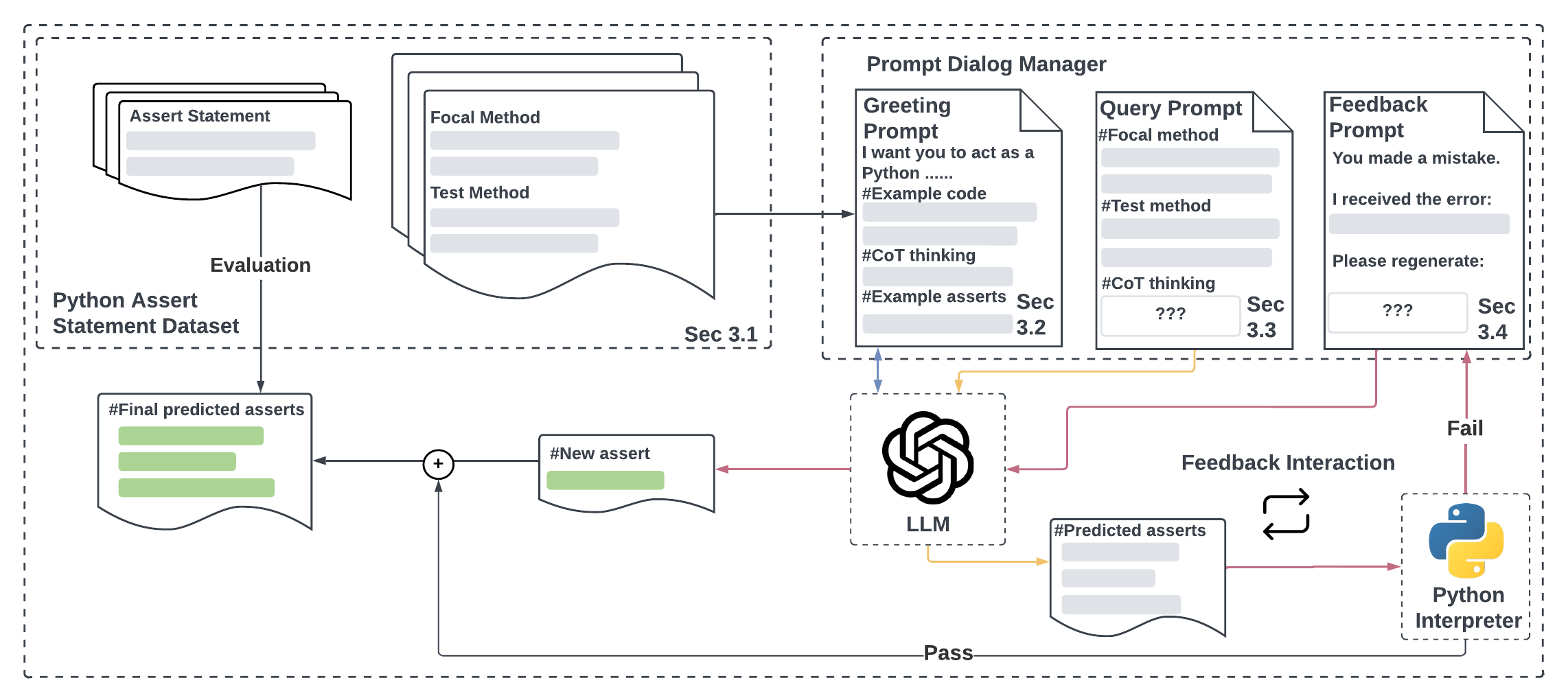}
	\caption{The overview of our approach \tool. }
	\label{fig:overview}
\end{figure*}

In this section, we introduce \tool, a system that engages the LLM as a role of an experienced unit test expert, and constructs prompts to initiate multiple conversation with the LLM to automatically generate the assert statements based on a given Python unit test case. Fig~\ref{fig:overview} illustrates the system overview of \tool. 
\han{Given a Python project, \tool analyses the project and creates data entries that include the focal methods, test methods, and corresponding assert statements. 
Then, \tool starts to communicate with the LLM, which involves three phases. First, we create a greeting prompt to define the task and provide one sample test case, offering LLM additional context (\ref{sec:greetingPrompt})}. In the second phase, we incorporate the target test case into the query and wait for the initial round of assert predictions (\ref{sec:queryPrompt}). In the third phase, we execute the predicted assert statements using a Python interpreter and, if any error messages arise, input these messages back into the LLM to request corrections to its predictions (\ref{sec:feedbackPrompt}). By interacting with the LLM, \tool collects the automatically generated assert statements and completes the unit test case.

\begin{figure}[!ht]
\begin{minted}{python}
# Focal Method
def process_spider_input(self, response, spider):
    meta = response.meta
    if 'handle_httpstatus' in meta:
    ......
    raise HttpError(response, 'Ignoring non-200 response')

# Unit Test
def test_spider_override_settings(self):
    self.spider.handle_httpstatus = [404]
    <AssertPlaceholder1>
    <AssertPlaceholder2>

# Assert Statements
self.assertIsNone(self.mw.process_spider_input(self.res404, self.spider))
self.assertRaises(HttpError, self.mw.process_spider_input, self.res402, self.spider)
\end{minted}
\caption{Example code after the data extraction.\protect\footnotemark}
\label{fig:tapSample}
\end{figure}
\footnotetext{\han{Code Sampled from: \url{https://github.com/scrapy/scrapy}}}

\subsection{Project Analysis}
\label{sec:methodologyDataExtraction} 
Our goal is to generate assert statements in Python projects. As of March 2023, Python has been acknowledged as the most popular programming language by the TIOBE Programming Community Index~\cite{web:languageRanking}; however, previous works have not focused on the assert generation area in Python. \han{Therefore, \tool begins by analysing the projects and preparing the data entries for the prompts construction. In Fig~\ref{fig:tapSample}, we provide an example of how a data entry extracted from a project is presented. A data entry consists of three main parts, the focal method, unit test, and assert statements. }

\han{Within a given project, \tool initially identifies all unit test files. This identification is based on the naming conventions of Python unit tests, as discussed in Section~\ref{sec:background1}, checking for specific patterns in the file names. Upon locating the files, \tool then examines their contents to verify the use of testing libraries and the completeness of the tests.
}

Next, \tool starts to identify the focal method of the giving test method. Following the focal method extraction for Java projects~\cite{watson2020learning, qusef2010recovering}, we identify the last method call from the same library or project before the assert statements as the focal method. In some cases, the method calls are embedded inside the assert statements as parameters, for example, the $process\_spider\_input$ in Fig~\ref{fig:tapSample}. This approach might be considered controversial, as the parameter to be predicted is already present in the prompts. However, in real-world scenarios, developers typically have a clear understanding of the focal method. Consequently, we believe that incorporating the focal method information in the prompts closely aligns with developers' actual workflow and context, providing a more realistic and practical representation of the problem at hand. \han{\tool retained the original comments within both the focal methods and the unit tests, as these help the language model understand the functionality of the functions. Global variables defined in the same file were also included.}

Finally, \tool extracts the assert statements from the unit test and use $\langle AssertPlaceholder\rangle$ to replace the assert statements within the unit test code. To differentiate assert statements in the same test case, \tool assigns incremental numbers to each of the statements. 
The detailed implementation is discussed in Section~\ref{sec:datasetCollection}. 

\begin{figure*}
    \centering
	\includegraphics[width=1\textwidth]{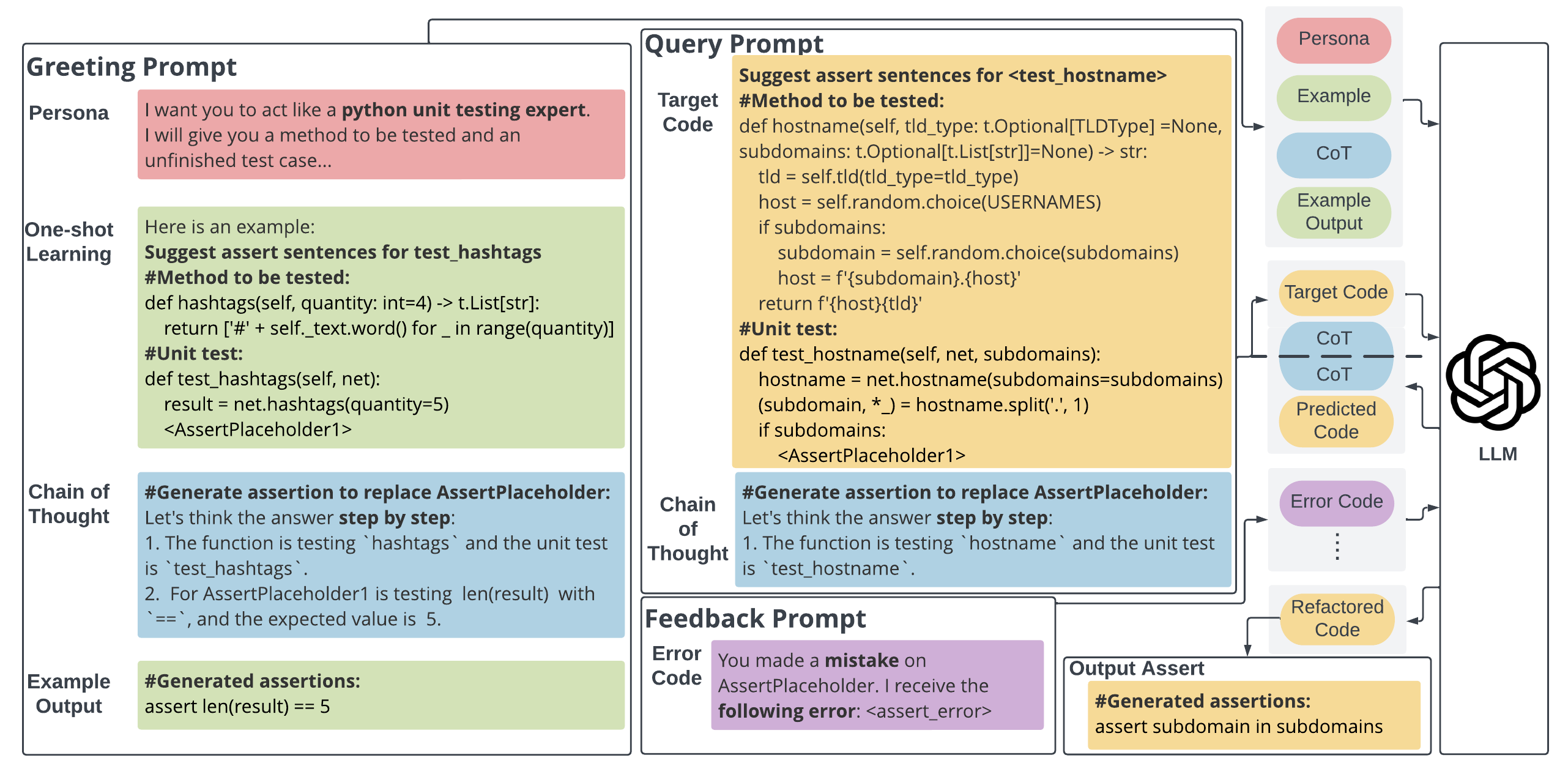}
	\caption{Prompt templates and sample responses.\protect\footnotemark}
	\label{fig:samplePrompt}
\end{figure*} 

\footnotetext{\han{Code sampled from: \url{https://github.com/lk-geimfari/mimesis}}}

\subsection{Greeting Prompt}
\label{sec:greetingPrompt}
The greeting prompt is the first prompt we sent to the LLM. As shown in Fig~\ref{fig:samplePrompt}, it includes a persona message, a sample code for one-shot learning, a Chain of Thoughts (CoT) prompting, and the corresponding assert statement as the output.

\textbf{Persona-based Prompting.} The persona-based prompting approach assigns a persona or role related to the task given to the LLM~\cite{white2023prompt}. By adopting a persona or role, the LLM can align its responses with the expertise, knowledge, and context associated with the given persona. This approach provides a clear framework and perspective for the LLM's response generation, ensuring that the output is more relevant and accurate for a specific task. In the official OpenAI documents, they also ask users to have a system message to help set the behaviour of the assistant~\cite{web:gpt}. 
In this study, we instruct the LLM to assume the role of a unit testing expert to assist in completing the unfinished test method by generating assert statements. The first sentence introduces the LLM to its assigned role. Then, inspired by a popular open-source repository that features a collection of valuable prompt examples~\cite{web:awesomprompts}, we include a task description paragraph to the prompt. The description details the input, steps to generate the assert statements, whether or not to use a certain testing framework, and the expected output, effectively guiding the LLM in producing the desired results.

\textbf{One-shot Learning.}
The LLM, such as GPT-3, can achieve state-of-the-art performance on a wide variety of tasks using only few-shot prompts, in which examples of solved tasks (shots) are included in the prompt. Our approach adopts the one-shot learning prompt to generate assert statements for Python projects, enabling the LLM to learn from a single example and adapt its response generation accordingly. This method considers Python's unique language characteristics and assumes that the writing style remains consistent across the same project. 
Therefore, while retrieving this one sample, we only look for the test method that is within the same test class of the target test method as the sample. 
If there is only one test method in a class, we use the other test method within the same project as the sample input.
When more than two test methods are present in a class, we compute the cosine similarity of the method names and choose the one with the highest score.  After locating the sample code, we parse it into a format suitable for constructing the Greeting Prompt, as illustrated in Fig~\ref{fig:samplePrompt}, and generate the corresponding CoT process. 
\han{We carried out a small-scale experiment showing that the performance of one-shot and few-shots prompting (6 shots) in our task does not show significant differences. In fact, one-shot prompting achieved higher accuracy (69.8\% over 58.1\%), likely due to less disruption in the prompts. Given the comparable performance and lower cost, we adopted one-shot prompting in \tool.}

\textbf{Chain-of-thought Prompting.} Recent research~\cite{wei2022chain, zhang2022automatic, kojima2022large} in prompt design has demonstrated that for complex tasks requiring multi-step reasoning, incorporating a rationale or key phrase can significantly enhance the performance of LLMs. The approach, called Chain-of-thought (CoT) prompting, is effective in guiding LLMs toward desired outputs. 
In our work, we implement CoT prompting strategies by incorporating a guiding phrase followed by rational steps for generating assert statements. This approach aims to leverage the advantages of both strategies to enhance the LLM's ability to produce accurate results. To develop the manual instruction steps, we investigated the process of writing unit tests and assert statements~\cite{web:howunittest, aniche2013asserts}. Additionally, we posed this question (i.e., What are the steps to write assertions for an incomplete unit test when provided with its focal method?) to the LLM model and combined the results to finalize the steps. The first step is to identify the test method and the focal method. Next, developers need to determine the parameters being tested and the assert type, e.g., asserting the parameter value, type/instance, or structural information. The final step involves reasoning the assert value and generating the corresponding statement.

\subsection{Query Prompt}
\label{sec:queryPrompt}
Following the submission of the greeting prompt, we introduce the actual query prompt containing the target test cases. The prompt follows a structure similar to the sample code in the greeting prompt but remains incomplete. We also include a brief description in the CoT section to remind the LLM model of the task, generating assertions to replace the assert placeholders, and request to start the CoT process from the first step.
The LLM then responds to \tool and produces the first round of predicted assert statements.


\subsection{Feedback Interaction and Prompt} 
\label{sec:feedbackPrompt}
Despite the LLM's ability to generate accurate assert statements, particularly with the CoT prompting which enables it to understand the testing purpose (derived from the test method and focal method names) and locate the parameter to be tested, there are instances where the LLM can not generate the correct value for the target parameters. \han{For example, in situations where the target parameter value relies on complex math calculations (e.g., deep learning models) or is influenced by external factors not immediately evident from the test method or focal method (e.g., parameters defined outside the files or obtained through multiple function calls, self-defined util functions), the LLM may fail to predict the correct value. }

To overcome the limitations, we introduce a feedback interaction that allows the LLMs to learn from execution errors in value generation, inspired by prior research~\cite{ni2023lever,daniel2009reassert, zamprogno2022dynamic}. As shown in Fig~\ref{fig:overview}, this process involves executing the initially generated assert statements using the Pytest library~\cite{web:pytest}.
If the assert passed, the LLM has identified the correct value for the tested parameter. \han{In cases where execution fails, \tool utilises regular expressions to automatically extract key information about the expected and actual values from the error message. \tool then incorporates the error message into the feedback prompt, ensuring that the LLM is aware of its previous mistake and can use this information to adjust its response generation accordingly for the specific assert placeholder.} Fig~\ref{fig:samplePrompt} shows a sample piece of the feedback prompt. Through this feedback interaction, the LLM can progressively generate precise values for target parameters. After all interactions, \tool collects the second-round predictions and saves them for evaluation.


\section{Evaluation}
\label{sec:evaluation}
To verify the performance of \tool, we conduct experiments to answer the following research questions:
\begin{itemize}
    \item \textbf{RQ1:} \han{How is the performance of \tool on LLMs}?
    \item \textbf{RQ2:} How do different prompt designs influence prediction performance?
    \item \textbf{RQ3:} How does the \tool perform in assert generation compare with the state-of-the-art approaches?
        \item \textbf{RQ4:} What are the cases generated by \tool that failed to accurately match the original test cases?
\end{itemize}


\subsection{Dataset}
\label{sec:datasetCollection}

\han{In our research, we assembled the first dataset for Python assert generation, by examining datasets from the recent popular Python research tools like Pynguin~\cite{lukasczyk2022pynguin}, BugsInPy~\cite{widyasari2020bugsinpy}, and CODAMOSA~\cite{lemieux2023codamosa}. Considering that CODAMOSA utilizes data from Pynguin and BugsInPy, we focused only on these two sources. To avoid the data leakage i.e., testing data being used for model training, we only included project files that were newly added after July 2021, which corresponds to the training data cut-off date for the GPT models~\cite{web:gpt}. 
After the removal, we identified 9 projects from the earlier works, with 7 of them using Python assert keyword and 2 employing testing libraries. To increase variety and robustness, we employed the crawling process of ATLAS~\cite{watson2020learning}, the first research work in the automatic assert generation field, identifying popular Python projects (stars and forks greater than 30) that also used testing libraries by Github APIs~\cite{web:githubAPI}. Combining the 5 most recently updated projects with the initial 9, we created a balanced dataset of 14 projects. These projects are equally distributed between those using the built-in assert keyword and those utilizing testing libraries such as unittest and numpy.testing.
}


\begin{table}[]
\resizebox{0.8\textwidth}{!}{
\begin{tabular}{llllll}
\hline
\textbf{Project}         & \textbf{Revision} & \begin{tabular}[c]{@{}l@{}}\# \textbf{Single Assert} \\ \textbf{Test Method}\end{tabular} & \begin{tabular}[c]{@{}l@{}}\# \textbf{Multi-assert}\\ \textbf{Test Method}\end{tabular} & \begin{tabular}[c]{@{}l@{}}\# \textbf{Total Test} \\ \textbf{Method}\end{tabular} & \textbf{Src}                                       \\     \hline  
dataclass-json   & 23e5eec  &                          4                                               &         0                                                              &                  4                                            & Pyn                                                   \\\rowcolor{gray!25}
mimesis          & 978d52c  &                         122                                              &                                          213                           &                   335                                           & Pyn                                                   \\
sanic            & 259e458  &                            169                                           &                             386                                          &                 555                                             & Pyn                                                   \\\rowcolor{gray!25}
black            & d9b8a64  &                          17                                              &                                       21                              &                     38                                         & \begin{tabular}[c]{@{}l@{}}Pyn \\ \& BIP\end{tabular} \\
ansible          & 367d45f  &                                      109                                  &                                          217                             &                          326                                    & BIP                                                   \\\rowcolor{gray!25}
cookiecutter     & cf81d63  &                                52                                         &                                           97                            &                                   149                           & BIP                                                   \\
thefuck          & ceeaeab  &                                    5                                     &                                                    6                   &                                           11                   & BIP                                                   \\\rowcolor{gray!25}
tornado$^\ast$    & 7186b86  &                                299                                         &                                     551                                  &                          850                                    & BIP                                                   \\
youtube-dl$^\ast$  & 0402710  &                          31                                               &                                         89                              &                     120                                         & BIP                                                   \\\rowcolor{gray!25}
scrapy$^\ast$     & 9cb757d  &                              222                                           &                                      479                                 &                       701                                       & \tool                                  \\
keras$^\ast$     & c72e310  &                                48                                         &                                      106                                 &                                154                              & \tool                                  \\\rowcolor{gray!25}
detectron2$^\ast$ & d779ea6  &                                41                                         &                                           79                            &              120                                                & \tool                                  \\
fairseq$^\ast$   & 176cd93  &                      9                                                   &                                                     33                  & 42                                         & \tool                                  \\\rowcolor{gray!25}
mkdocs$^\ast$     & 56b235a  &                                      67                                   &                                             224                          &                                291                              & \tool       \\ \hline
\textbf{Total} & - & 1,195 & 2,501 & 3,696 & -  \\
	\hline                          
\end{tabular}
}
\caption{Characteristics of the Python repositories used in the evaluation.\protect\footnotemark} 
\label{tab:DatasetTable}
\end{table}

Given a project, \tool identifies and extracts unit test suites, we wrote automated scripts that could parse the selected repositories, traverse the directory structure, and pinpoint the test files according to the patterns we discussed in Section~\ref{sec:methodologyDataExtraction}. Then, we excluded test methods created before July 2021 to match with the cut-off date. Following the approach used by CODAMOSA, we employed pipreqs~\cite{web:pipreqs} to identify project dependencies and configured the testing environment accordingly. We down-sampled some methods that necessitated additional setup procedures. \han{In total, the dataset consists of 3,696 test methods and 1,298 focal methods (2.85 test methods per focal methods on average).} Table~\ref{tab:DatasetTable} provides a detailed overview of the projects included in our dataset. 

\footnotetext{$^\ast$ denotes the project uses testing libraries. \# Test Method is the number of test methods within the project. Src is the project source, where Pyn is for Pynghuin, BIP is for BugsInPy, and \tool is from this study.}

To match the focal method with the test case, \tool traverses through the Abstract Syntax Tree (AST)~\cite{baxter1998clone} of the unit test code using the $NodeVisitor$ class, and identifies the last method call from the same project before or within the assert statements.
In Section~\ref{sec:background}, we discussed two prevalent approaches for writing assert statements. When using the build-in assert, the assert statements consistently begin with the $assert$ keyword, while when employing popular testing libraries, the unit test methods generally inherit the test class provided by the libraries, and their assert statements start with $self.assert{FunctionName}$. In our extraction script, we combine the use of AST and regular expressions to effectively extract assert statements from the original source code. We executed the scripts across all projects and constructed the dataset for evaluation. The dataset shall also benefit relevant studies within the community.

In our testing, the LLM is capable of generating assertions corresponding to the number of placeholders provided. But in the evaluation, we aligned the number of assert placeholders with the original test cases, maintaining a one-to-one correspondence. This process is reflective of typical developer practices when crafting test methods, where developers often write assertions sequentially, having prior knowledge of the required quantity and understanding when the test case is complete~\cite{web:onlineTutorial, web:onlineTutoria2, web:onlineTutoria3}. By limiting the number of assert placeholders, we aimed to assess whether \tool could generate assertions efficiently.

\subsection{Evaluation Metrics}
We use the following metrics to evaluate the generated assert statements by the LLM.

\textbf{Accurate Match (AM \%).} The Accurate Match metric measures the functional equivalence of the generated asserts to the original asserts. In addition to exact matches, we extracted assert statement types from Python testing libraries and identified 34 groups of assert pairs that, while not identical, perform the same function. This concept is inspired by suboptimal test smells~\cite{wang2021pynose}. Fig~\ref{fig:suboptimal} provides some examples of these equivalent assert pairs.
For example, in the first pair of assertions, they are functionally same, but the predicted one is preferred as it conveys the intention more explicitly, making the code more readable and maintaining consistency with standard testing practices.

\begin{figure}[!ht]
\begin{minted}{python}
self.assertLen(js, 3) #predicted
self.assertEqual(len(js), 3) #original

self.assertIsNone(failure) #predicted
self.assertTrue(failure is None) #original

self.assertRaises(ValueError, text_response.json) #predicted
with self.assertRaises(ValueError):
    text_response.json   #original
\end{minted}
\caption{Example of suboptimal assert match.}
\label{fig:suboptimal}
\end{figure}

\textbf{LCS (\%).}  We measure the longest common subsequence between the predicted and original assert, calculating the percentage as in Nashid et al.~\cite{nashidretrieval}, by dividing the subsequence length by the original assert length.

\textbf{Edit Distance (ED).} The edit distance is a metric to measure the similarity between two strings by calculating the minimum number of single-character edits (e.g., insertions, deletions, or substitutions) required to transform one into the other. The metric has been widely used to evaluate code-related tasks~\cite{chakraborty2021multi, ding2020patching, nashidretrieval}.

\textbf{Assert Method Match (AMM \%).} \han{The AMM is a metric to verify whether the predicted assert statement type aligns with the original one. For example, if CLAP predicts an assert statement as \textit{assertIsInstance(a, List)}, but the actual assert statement is \textit{assertIsInstance(a, pd.Array)}, this is still classified as a match under the AMM metric as they both want to check the instance of parameter a. The calculation of AMM is represented by the following formula:}

\[AMM = \frac{1}{N} \sum_{i=1}^N \mathbb{I}\left( AssertMethodPredict_i= AssertMethodOriginal_i\right)\]where $\mathbb{I}$
$(AssertMethodPredict_i= AssertMethodOriginal_i)$ is an indicator. When the predicted assert method is equal to the original assert method, $\mathbb{I}=1$, otherwise, 0.

\han{We use this metric because, the assert statements in Python come in various types, e.g., assertIn, assertIsNone, assertAlmostEqual, and assertGreaterEqual. The AMM metric quantifies the precision of CLAP in predicting the appropriate assert method, thereby demonstrating its ability in predicting relevant assert predictions based on the current context. Note that the metric is specifically used to evaluate projects employing testing libraries, as the assert keyword does not distinguish between different assert methods.}

\begin{table}[]
\centering
\resizebox{0.7\textwidth}{!}{
\begin{tabular}{l|l|lllll}
	\hline  
Framework                                                                                & Assert Num.                    & Model         & AM(\%)        & LCS(\%)       & ED            & AMM(\%)               \\ \hline
\multirow{4}{*}{\begin{tabular}[c]{@{}l@{}}Python\\ Assert\\ Keyword\end{tabular}}      & \multirow{2}{*}{Single Assert} & text-davinci  & \textbf{64.6}   & \textbf{81.1} & \textbf{11.3} & \multicolumn{1}{c}{-} \\
                                                                                         &                                & \cellcolor{gray!25}gpt-3.5-turbo & \cellcolor{gray!25}55.2          & \cellcolor{gray!25}77.1          & \cellcolor{gray!25}15.9          & \multicolumn{1}{c}{\cellcolor{gray!25}-} \\ \cline{2-2}
                                                                                         & \multirow{2}{*}{Multi Assert}  & text-davinci  & \textbf{51}   & \textbf{74.1} & \textbf{11.7} & \multicolumn{1}{c}{-} \\
                                                                                         &                                & \cellcolor{gray!25}gpt-3.5-turbo & \cellcolor{gray!25}46.3          & \cellcolor{gray!25}69.4          & \cellcolor{gray!25}11.5          & \multicolumn{1}{c}{\cellcolor{gray!25}-} \\ \hline
\multirow{4}{*}{\begin{tabular}[c]{@{}l@{}}Python\\ Testing\\ Library\end{tabular}} & \multirow{2}{*}{Single Assert} & text-davinci  & \textbf{66.7} & \textbf{81.8} & 13.1          & \textbf{84.8}         \\
                                                                                         &                                & \cellcolor{gray!25}gpt-3.5-turbo & \cellcolor{gray!25}57.3          & \cellcolor{gray!25}76            & \cellcolor{gray!25}\textbf{12.7} & \cellcolor{gray!25}77.8                  \\ \cline{2-2}
                                                                                         & \multirow{2}{*}{Multi Assert}  & text-davinci  & \textbf{68.7} & \textbf{84.4} & \textbf{8.9 }         & \textbf{83.3}           \\
                                                                                         &                                & \cellcolor{gray!25}gpt-3.5-turbo & \cellcolor{gray!25}51            & \cellcolor{gray!25}69            & \cellcolor{gray!25}10.7 & \cellcolor{gray!25}72.1                  \\ \hline
\multirow{2}{*}{Average}                                                                   &                                & text-davinci  & \textbf{62}   & \textbf{80.6} & \textbf{10.6}         & \textbf{83.7}           \\
                                                                                         &                                & \cellcolor{gray!25}gpt-3.5-turbo & \cellcolor{gray!25}51            & \cellcolor{gray!25}71.1          & \cellcolor{gray!25}11.8 & \cellcolor{gray!25}73.9               \\	\hline    
\end{tabular}
}
\caption{Detail statistics of \tool with \textit{text-davinci} model and \textit{gpt-3.5-turbo} model}
\label{tab:CompareModel}
\end{table}

\subsection{Performance of \tool (RQ1)}

\textbf{Experiment setup.} In this section, we evaluate the prompt design across two LLM models from OpenAI~\cite{web:gpt}: \textit{text-davinci-003}(\textit{text-davinci}) and \textit{gpt-3.5-turbo}. Inspired by previous works~\cite{paranjape2023art} and a small-scale pilot study, we set the temperature parameter to 0.3, striking a balance between allowing some degree of creativity in the model's responses while ensuring that the predictions remain well-defined. We understand the \textit{code-davinci} is trained on more code repositories, but it was deprecated~\cite{web:gpt}, we have discussed this threat in Section~\ref{sec:threat}. All requests were executed through Python scripts utilizing the official OpenAI Python APIs.

\begin{table}[]
\resizebox{0.95\textwidth}{!}{
\begin{tabular}{l|llll|llll|llll}
\hline
\multirow{2}{*}{Project} & \multicolumn{4}{c|}{Single Assert} & \multicolumn{4}{c}{Multiple Asserts} & \multicolumn{4}{c}{Single + Multi Asserts}\\
                         & AM(\%)     & LCS(\%)     & ED      & AMM(\%)   & AM(\%)      & LCS(\%)     & ED      & AMM(\%)    & AM(\%)      & LCS(\%)     & ED      & AMM(\%) \\ \hline
dataclass-json           & 100    & 100     & 0       & -     & -     & -     & -       & -  & 100     & 100     & 0       & -     \\\rowcolor{gray!25}
mimesis                  & 65.3     & 79.7    & 10.5    & -     & 45.2      & 70.0    & 11.2      & -  & 51.2      & 73.3    & 11      & -    \\
sanic                    & 55.6     & 76    & 15.9    & -     & 48.7      & 75.2    & 10.2    & -   & 50      & 75.4    & 11.3    & -   \\\rowcolor{gray!25}
black                    & 94.1     & 95.4    & 1.3     & -     & 61.5      & 79.8    & 6.8     & -  & 81.5      & 89.5    & 3.4     & -     \\
ansible                  & 71.1     & 87.1    & 6.8     & -     & 54.7      & 72.6    & 15.0    & -   & 58.5      & 75.6    & 13.1    & -   \\\rowcolor{gray!25}
cookiecutter             & 61.5     & 75.7    & 15      & -     & 60.4      & 82.3    & 14.6    & -   & 60.7      & 80.5    & 14.7    & -   \\
thefuck                  & 79.2     & 96.5    & 2.5     & -     & 50.0      & 63.0    & 15.0     & -   & 76.9      & 93.9    & 3.4     & -   \\\rowcolor{gray!25}
tornado                  & 72.3     & 84.3    & 9.2    & 86.7    & 62.8      & 80.3      & 9.6    & 79.0   & 66.2      & 81.7      & 9.5    & 81.7   \\
youtube-dl               & 83.3     & 90.1      & 7.5     & 91.7    & 66.7      & 88.0      & 7.3    & 87.5   & 70.8      & 88.5      & 7.3    & 88.5  \\\rowcolor{gray!25}
scrapy                   & 52.6     & 78.2    & 22      & 82.7    & 65.6      & 83.1    & 10.4    & 78.6  & 61.4      & 82.1    & 12.7    & 79.3   \\
keras                    & 59.4     & 76.8    & 11.3     & 71.9   & 67.3      & 80.8      & 11.5    & 86.7   & 65.5      & 79.8      & 11.4    & 82.7  \\\rowcolor{gray!25}
detectron2               & 70.6     & 84      & 5.3     & 91.2    & 47.1      & 66.7    & 14.5    & 80  & 48.1      & 72.4    & 11.5    & 83.7   \\
fairseq                  & 60     & 83.4    & 4.6     & 80   & 55.8      & 81.5    & 10.1     & 92.8  & 56.2      & 81.7    & 9.2     & 87.5  \\\rowcolor{gray!25}
mkdocs                   & 59.6     & 79.2      & 12.9     & 82.5    & 83.8      & 91.5    & 5.6    & 91.0  & 81.7      & 90.4    & 6.3    & 90.3  \\ \hline
\end{tabular}
}
\caption{Detail statistics of \tool per project.}
\label{tab:ProjectResult}
\end{table}

\textbf{Result.} Table~\ref{tab:CompareModel} shows the results of the assert generation tasks for the two LLMs utilized by \tool. We can observe that for the Python assert keyword, \textit{text-davinci} outperforms \textit{gpt-3.5-turbo} in both single and multiple assert generation scenarios, achieving 64.6\% and 51\% accurate match respectively. The prediction on the single assert from \textit{text-davinci} has nearly 10\% more accurate assert than the \textit{gpt-3.5-turbo} model. For testing libraries, \textit{text-davinci} till achieves better performance in most metrics. Conversely, \textit{gpt-3.5-turbo} exhibits a lower ED for single assert generation.
Overall, we found the \textit{text-davinci-003} model generally outperforms the \textit{gpt-3.5-turbo} model in generating accurate assert statements. 
Although \textit{gpt-3.5-turbo} is designed for chat-like interactions, which aligns with our prompt design, it appears to be more cautious and hesitant to avoid making errors. For instance, in some cases, it fails to predict an assert statement because it seeks additional information about method calls beyond the focal methods, particularly for complex test input and multi-assert generation. In contrast, \textit{text-davinci} seems capable of learning about other method calls based on their names and parameters, enabling more accurate predictions. Furthermore, the turbo architecture of \textit{gpt-3.5-turbo} may impact its understanding of longer assertions, resulting in reduced accuracy.

To explore more details, we further analysed the \textit{text-davinci} results at the project level. Table~\ref{tab:ProjectResult} displays the statistics across various projects. \textit{Dataclass-json} achieves a 100\% AM, because it only has four assert methods. \textit{Detectron2} has the lowest AM overall, possibly due to its complex DL model structure. 
\textit{Sanic} and \textit{Scrapy} also show lower AMs, possibly because their web-related nature demands specific testing environments and assert parameters. For 11 out of 14 projects, the AM decreases from the single assert generation to all assert generation. However, \textit{Scrapy}, \textit{Keras}, and \textit{Mkdocs} show an increased AM for all test methods, suggesting higher accuracy in predicting multi-assertions. This can be attributed to these three projects having some test cases with similar structural assert statements, therefore increasing the AM overall.

\begin{table*}[]
    \centering
    \resizebox{1\textwidth}{!}{
\begin{tabular}{l|l|llll|llll|llll|llll}
\hline
\multicolumn{1}{c|}{}                                                              & \multicolumn{1}{c|}{}                                                                         & \multicolumn{4}{c|}{\textbf{CLAP}}                                                                                                                                    & \multicolumn{4}{c|}{\textbf{CLAP\_noPersona}}                                                                                               & \multicolumn{4}{c|}{\textbf{CLAP\_noCoT}}                                                                                                                             & \multicolumn{4}{c}{\textbf{CLAP\_noFDBK}}                                                                                                  \\
\multicolumn{1}{c|}{\multirow{-2}{*}{Framework}}                                   & \multicolumn{1}{c|}{\multirow{-2}{*}{\begin{tabular}[c]{@{}c@{}}Assert \\ Num.\end{tabular}}} & AM(\%)                                & LCS(\%)                               & ED                                   & AMM(\%)                                        & AM(\%)                       & LCS(\%)                      & ED                           & AMM(\%)                                        & AM(\%)                                & LCS(\%)                               & ED                                   & AMM(\%)                                        & AM(\%)                       & LCS(\%)                      & ED                           & AMM(\%)                                       \\ \hline
                                                                                   & Single Assert                                                                                 & \textbf{64.6}                         & \textbf{81.1}                         & \textbf{11.3}                        & \multicolumn{1}{c|}{-}                         & 62.6                         & 79.9                         & 11.4                         & \multicolumn{1}{c|}{-}                         & 60.6                                  & 78.8                                  & 12.1                                 & \multicolumn{1}{c|}{-}                         & 53.8                         & 76.6                         & 13.7                         & \multicolumn{1}{c}{-}                         \\
\multirow{-2}{*}{\begin{tabular}[c]{@{}l@{}}Python Assert\\ Keyword\end{tabular}}  & Multi Assert                                                                                  & \cellcolor{gray!25}51          & \cellcolor{gray!25}74.1          & \cellcolor{gray!25}11.7         & \multicolumn{1}{c|}{\cellcolor{gray!25}-} & \cellcolor{gray!25}47.7 & \cellcolor{gray!25}74.2 & \cellcolor{gray!25}12.2 & \multicolumn{1}{c|}{\cellcolor{gray!25}-} & \cellcolor{gray!25}\textbf{52.3} & \cellcolor{gray!25}\textbf{76.4} & \cellcolor{gray!25}\textbf{9.9} & \multicolumn{1}{c|}{\cellcolor{gray!25}-} & \cellcolor{gray!25}46   & \cellcolor{gray!25}73.4 & \cellcolor{gray!25}12.1 & \multicolumn{1}{c}{\cellcolor{gray!25}-} \\ \hline
                                                                                   & Single Assert                                                                                 & \textbf{64.7}                         & \textbf{81.8}                         & \textbf{13.1}                        & \textbf{84.8}                                  & 62.8                         & 81.6                         & 13.9                         & 81.2                                           & 62.6                                  & 79.7                                  & 13.4                                 & 78.4                                           & 43.5                         & 71.9                         & 19                           & 74                                            \\
\multirow{-2}{*}{\begin{tabular}[c]{@{}l@{}}Python Testing\\ Library\end{tabular}} & Multi Assert                                                                                  & \cellcolor{gray!25}\textbf{68.7} & \cellcolor{gray!25}\textbf{84.4} & \cellcolor{gray!25}\textbf{8.9} & \cellcolor{gray!25}83.3                   & \cellcolor{gray!25}60   & \cellcolor{gray!25}79.7 & \cellcolor{gray!25}11.4 & \cellcolor{gray!25}79                     & \cellcolor{gray!25}67.6          & \cellcolor{gray!25}84.3          & \cellcolor{gray!25}9            & \cellcolor{gray!25}\textbf{85.6}          & \cellcolor{gray!25}45.7 & \cellcolor{gray!25}71.9 & \cellcolor{gray!25}13.8 & \cellcolor{gray!25}77.7                  \\ \hline
Average                                                                            &                                                                                               & \textbf{62}                           & 80.6                                  & 10.6                                 & \textbf{83.7}                                  & 58.2                         & 78.6                         & 11.5                         & 80                                             & 61.5                                  & \textbf{80.7}                         & \textbf{10.2}                        & \textbf{83.7}                                  & 46.4                         & 73                           & 14.1                         & 77.4                                          \\ \hline
\end{tabular}}
\caption{Results for prompt design evaluation.}
\label{tab:promptEvaluation}

\end{table*}

\begin{table}[]
\resizebox{0.7\textwidth}{!}{
\begin{tabular}{l|llllllll}
\hline
             & \textbf{Equal} & \textbf{True} & \textbf{Is} & \textbf{False} & \textbf{Raises} & \textbf{In} & \textbf{IsInstance} & \textbf{Other} \\ \hline
CLAP\_noFDBK & 48.9           & 43.8          & 60          & 46.8           & 21.9            & 26.3        & 60                  & 31.9           \\ 
CLAP\_Feed   & 58.7           & 56.9          & 65          & 52.7           & 30.1            & 33.3        & 64.7                & 37             \\ \hline
\end{tabular}}
\caption{Detailed AM(\%) statistics of \tool for each assert type.}
\label{tab:assertType}
\end{table}

\subsection{Effectiveness of the Prompt Design (RQ2)}
\label{sec:rq2}

\textbf{Experiment setup.}
We assess the effectiveness of the prompt design with various input configurations. Existing research has confirmed that providing examples to the LLM can improve accuracy for different tasks~\cite{nashidretrieval, ahmed2022few, brown2020language}. Therefore, we evaluate the other three features within the \tool prompt design, which include persona, CoT, and feedback prompting. In the experiment, we remove the portions of the prompt that serve these features respectively and maintain all other settings. We then collect the results from different configurations and evaluate their performance.

\textbf{Result.}
Table~\ref{tab:promptEvaluation} presents the results of the prompt design evaluation.  Removing persona design leads to decreased performance, revealing the importance of persona prompts in generating accurate assert statements. The situation is different for the CoT prompt design validation. For single assert statements, \tool outperforms without CoT for both the Python assert keyword and testing libraries. However, multi-assert generation shows mixed results. For Python assert keyword tests, \texttt{CLAP\_noCoT} performs better; for Python testing libraries, \tool's prompt achieves higher AM and LCS.
Upon examining the multi-assert generation cases responsible for this difference, we identified the following reasons: 1) Some multi-assertion test cases test multiple parameters within a single unit test, which, although common, is considered poor practice for code quality~\cite{aniche2013asserts}. For instance, a test case might use 8 assertions to test 3 parameters. With the CoT prompt design, the LLM might mistakenly use extra assertions to test the type of parameter A rather than the value of parameter B, ultimately neglecting to test parameter B and resulting in lower AM and AMM. 2) Some test cases are too long and come close to the token limitation of the GPT models, leading the LLM to generate assert statements that might not conform to the desired format. This makes it challenging for \tool to identify the predictions. Nevertheless, the average AM of \tool remains higher than the design without CoT. Future work should focus on improving the prompt design for multi-assert statements to address the issues mentioned above.

In general, we observe that the most significant difference across the four metrics lies between \texttt{CLAP\_noFDBK} and \tool, indicating that the feedback interaction design has a substantial impact on performance. 
Various scenarios may cause the LLM to struggle in predicting the correct value, such as global parameters or functions defined outside the focal method, asserts requiring mathematical calculations, or focal methods with dependencies on other functions. Feedback interaction can assist the LLM in learning from error messages and correcting incorrect predictions. To gain deeper insights into which assert types benefit the most, we provide a breakdown of AM statistics for different assert types in Table~\ref{tab:assertType}. The assert types \textit{Is} and \textit{IsInstance} achieve the highest accuracy, possibly due to the complexity of predicting whether objects are the same and determining the object's type is low. The feedback interaction contributes to the most significant accuracy increase for \textit{Equal} and \textit{True} types, which is because the feedback interaction gives the LLM a better understanding to the expected values. For example, if the \textit{assertEqual} predicted value is incorrect in the first round, the LLM can comprehend the correct value based on the feedback prompt and adjust its prediction accordingly.

Overall, all three prompt components contribute positively to assert generation, with feedback interaction having the most significant impact on performance, while the CoT prompt design needs to be improved for the multi-assert generation.

\subsection{Comparison with State-of-the-Art Approaches (RQ3)}
\textbf{Baseline setup.}
We evaluate the performance of \tool by comparing it to other state-of-the-art assert generation approaches. ATLAS~\cite{watson2020learning} is the first research work in this field and has demonstrated strong results by adopting a learning-based method. 
We also fine-tune CODET5~\cite{wang2021codet5}, a pre-trained model that is capable of various code-related tasks such as code completion, employing it as one of the baseline models in our study.
CEDAR~\cite{nashidretrieval} is another approach that leverages the capabilities of LLMs in combination with information retrieval techniques. Since all prior approaches have exclusively focused on single assert generation, we also evaluate them with the single assert dataset. For ATLAS and CodeT5, we followed their paper instructions to set up the models and validated the result with five-fold cross validation~\cite{web:crossvalidation}. For CEDAR, we strictly followed their codebase to use the BM25~\cite{robertson2009probabilistic} for the information retrieval and the rest of their code to do the data preprocessing and communicate with OpenAI API. 
To minimize any potential bias that could be introduced through the IR approach, we adopted the concept of Leave-One-Out~\cite{hastie2009elements}, which is known to reduce bias while optimizing prediction results in comparison to using a single test set. Specifically, when predicting an assert statement for a given test case, we employ the remaining test cases as the training set for IR, subsequently calculating the evaluation metrics across the entire dataset. Note that since the CODEX model was deprecated, we use the \textit{text-davinci-003} as the LLM to predict. We acknowledge there are other approaches in assertion generations~\cite{mastropaolo2022using, mastropaolo2021studying, dinella2022toga}. However, they have some specific settings to Java which presents a challenge to the replication of their method in this study.

\begin{table}[]

\centering
\begin{tabular}{l|llll}
\hline
\textbf{Approach}    & \textbf{AM(\%)} & \textbf{LCS(\%)} & \textbf{ED} & \textbf{AMM(\%)} \\ \hline
\texttt{ATLAS}            &       7.8         &       38.7         &      31.3       &        67         \\
\texttt{CodeT5}                & 25.1          & 60            & 39        & 80.4            \\
\texttt{CEDAR}                & 54.7            & 76.6             & 16.3        & 81.5             \\
\tool & \textbf{64.7}     & \textbf{81.5}     & \textbf{12.4}     & \textbf{84.8}            \\ \hline
\end{tabular}
\caption{Compare the performance of \tool with other state-of-the-art approaches.}
\label{tab:baselineResult}
\end{table}

\textbf{Result.} Table~\ref{tab:baselineResult} presents a performance comparison of \tool with ATLAS, CodeT5, and CEDAR. In all metrics, \tool outperforms the other approaches. Both \tool and CEDAR employ LLM to generate assert statements, leading to improvements in AM. Nevertheless, the difference in AMM is relatively small, suggesting that learning-based approach can identify the correct type of assert but fail to accurately capture the specific values or parameters within the assert statement. For example, given a statement $self.assertEqual(out, reqs)$, the output from ATLAS is $self.assert$ $Equal(self, data)$, in which the two parameters within the assert statement do not match any elements from the original one. Sometimes the prediction from ATLAS does not follow the Python syntax, as the learning-based approach relies on patterns observed during training, which may not always adhere to the language's syntax rules. 
CodeT5 has a similar AMM when compared with CEDAR and \tool, but falls short in other metrics, particularly ED. We observed instances where CodeT5 failed to terminate its output, resulting in the generation of extraneous and irrelevant text.

CEDAR achieves 54.7\% in AM, which demonstrates its capability to generate meaningful assert statements. However, when compared to \tool, there is still room for improvement. 
The use of IR includes more shots into the prompt but compared with \tool, some shots may be inconsistent with the context, affecting the LLM's prediction.  For instance, for less common assert types like $assertIsInstance$, \tool can predict the correct assertions, while CEDAR might be influenced by other samples and predict the value using $assertEqual$ or $assertTrue$, which are more widely-used assert types.
 Besides, the incorporation of CoT and feedback in the prompt design enables \tool to better predict assert statement values accurately, especially for the complex value structure. For example, given the assert statement $self.assertEqual(jsi.call\_function('x'),$ $ [20, 20, 30, 40, 50])$, \tool can identify the correct value with CoT's assistance even before entering the feedback interaction. 

\begin{figure}
    \centering
	\includegraphics[width=0.7\textwidth]{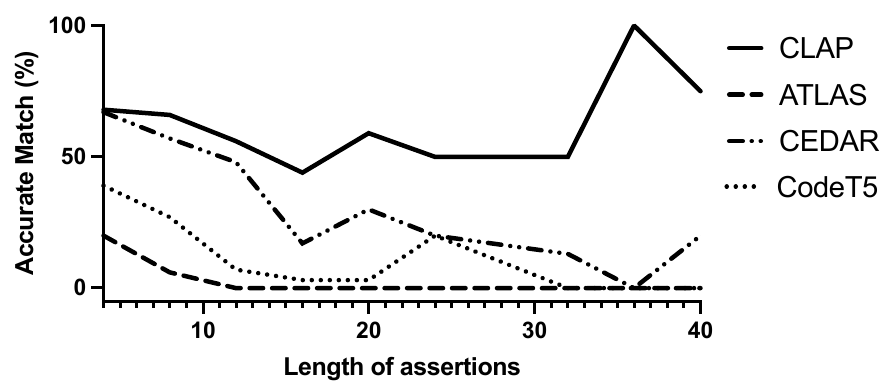}
	\caption{Length distribution of correct predicted assertions across \tool, CEDAR, ATLAS, and CodeT5.}
	\label{fig:length}
\end{figure}

 Fig~\ref{fig:length} shows the distribution of correct assertion lengths for the three approaches. It can be observed that, for shorter assertions, ranging up to 16, \tool is capable of generating better predictions in the majority of cases. This can be attributed to the prompt design that assists in comprehending the test input and enables adjustments to the predicted value if it is initially incorrect. For longer assertions, the performance fluctuates, but \tool still can predict the correct result for long assertions.
 Nonetheless, \tool demonstrates improvements in performance and generates more accurate assert statements with less context.

We conducted further examination of the AM assertions generated by \tool, specifically focusing on cases that are AM but not exact matches. Among the randomly selected 50 test cases, 10\% of them are due to the assert parameters are inverted, whereas the rest 90\% them are due to readability differences. One prevalent pattern in such cases involves substituting \textit{assertTrue} and \textit{assertFalse} with more descriptive assertions like \textit{assertEqual}, \textit{assertLen}, \textit{assertIsInstance}, and \textit{assertNone} (illustrated in Fig~\ref{fig:suboptimal}). Though functionally identical, the predicted assertions are preferred as they articulate the intention more explicitly, enhancing code readability and aligning with established testing practices, thereby improving the overall quality of the test code. Another pattern pertains to \textit{assertRaises} assertions, where readability can either improve or decline depending on the complexity of the code or developer preference. For instance, in the final example of Fig~\ref{fig:suboptimal}, the predicted assertion is a concise one-line explanation, in contrast to the more space-consuming original code. However, as the lines in the raise block expand, the one-line format commonly produced by \tool may become more challenging to interpret. The findings highlight \tool has the ability to increase the readability and maintain consistency with standard testing practices, demonstrating its potential to not only accurately replicate functionality but also enhance the overall quality of the test code.

\subsection{Analysis of the Non-AM Assertions (RQ4)}

\textbf{Motivation \& Experiment setup.} 

In this RQ, we conduct a qualitative analysis of the assert statements generated by \tool that are not AM to the original ones. Through a detailed examination, we can identify the strength or weaknesses of the current approach, and gain valuable insights into the complexities and challenges of automated assert generation. The assessment is conducted by the first two authors who are software engineer PhD students with substantial experience in Python development. We implement an open coding approach~\cite{seaman1999qualitative} to evaluate the generated assertions. 
A total of 150 test cases are randomly selected from across all non-AM test methods, satisfying a statistically significant sample size that ensures a 95\% confidence level and a 0.08 confidence interval~\cite{zar1999biostatistical}. 
Initially, the first 30 cases are jointly examined by the two authors. Subsequently, we independently label the remaining cases. In instances where the generated asserts are better than the original asserts, we submitted pull requests to the corresponding repositories.

\begin{figure}[!ht]
\begin{minted}{python}
#1
expected = 'something' # defined in the test case
self.assertEqual(output, expected) #predicted
self.assertEqual(output, 'something') #original
#2 
assert response.status == 200 #predicted
assert response.text == 'OK' #original
#3
self.assertEqual(result, [0]) # predicted
self.assertEqual(result, [1,4,7,8]) #original
#4
self.assertDictEqual(result, expected_result) #predicted
self.assertEqual(result['site_name'], 
                  expected_result['site_name']) #original
#5
assert 1 <= date <= 31 #predicted
assert (date >= 1) or (date <= 31) #original
\end{minted}
\caption{Example of some predictions that are non-accurate matches.}
\label{fig:discussion}
\end{figure}

\textbf{Result.}

Through the analysis, we identified cases that should be considered as correct predictions, but are challenging to detect using automated scripts. For the assertions that were predicted incorrectly, we also summarized three typical scenarios and investigated the underlying reasons behind these inaccuracies. 

22\% of the non-AM test cases in our randomly selected dataset should be considered correct predictions. 
For example, in Fig~\ref{fig:discussion} sample 1, the predicted assertions use the predefined parameter whereas the original one has a hard-coded string. In sample 2, the generated assert checks the response status, whereas the original developer checks the response text. The purpose of both is to check if the response is successful, but they assert different attributes of the HTTP response. The situation varies across different cases, but these examples illustrate that the non-AM test cases may represent alternative but valid ways of performing the assertion as well. 

We identified three categories of incorrect predictions in the assertions: value/attribute error (31\%), different assert rules (22\%), and unrelated (25\%). Firstly, the value/attribute error refers to the generated assertions inaccurately reflecting the values being asserted or asserting the wrong attribute of a parameter (sample 3 for example). Though the feedback interaction prompt is designed to help \tool with getting the correct value, for some complex setups, it may still fail to capture the precise value. 
Secondly, there are instances where predicted assertions enforce conditions that are either more or less stringent than the original. The fourth sample reveals a stricter predicted assertion, checking the complete equality of the entire dictionaries, as opposed to the original code's specific focus on the site\_name key.
Thirdly, the unrelated cases are where the generated assertions have no clear connection to the real tested parameters, which shows the \tool may have misapprehended the test case's objective and assessed the wrong parameters. The reasons could stem from the length of the test cases, obscured method calls, or other complexities.
In addition, although it's rare for popular open-source projects, we identified a bug in the original test code. In the last sample, the predicted assertion evaluates if date lies within the range of 1 to 31, offering an accurate condition than the original assertion's potentially flawed use of the 'or' operator.

In analyzing non-AM test cases, we found that 22\% should be considered correct predictions, representing valid alternatives to the original assertions. Among the incorrect predictions, 53\% were closely related to the tested parameters, falling into categories of value/attribute error or different assertion rules. These cases might require minor modifications to align with the original intention. The findings highlight that a significant portion of non-AM cases either align with correct predictions or maintain relevance, underscoring its practical potential.

Following the coding exercise, we conducted a repository-wise summary of the generated assertions that could improve code readability, eliminate test smells, or establish more accurate rules in contrast to the original statements (e.g., sample 5) from both the AM but not exact match cases and the non-AM test cases but should be considered correct predictions. We made submissions in the form of pull requests (PRs) to a total of 9 repositories, resulting in 181 proposed line changes. By the time of this paper's submission, 7 PRs had been approved or merged while the rest were still under review. We also received thankful comments from project maintainers expressed for our efforts in refining the test code and rectifying potential bugs. These validations from the real software development community highlight the practical utility and efficacy of \tool in enhancing the quality of unit test code.

\section{Discussion}
\label{sec:discussion}
\subsection{Generalisation across LLMs}
\han{As LLMs gain popularity, various new models are continually being developed by different companies and groups. To evaluate the effectiveness of CLAP's performance across a range of LLM models, we extended our evaluation beyond the initially tested models. Using the same experimental setup as RQ1, we included additional LLMs, specifically OpenAI's GPT-4~\cite{web:gpt4API} and Google’s Gemini 1.5 Pro~\cite{web:gemini}. Table~\ref{tab:discussion} presents the results from these models compared to \textit{text-davinci-003}, which achieved the highest performance in RQ1.}

\begin{table}[]
\resizebox{0.7\textwidth}{!}{
\begin{tabular}{l|ccc}
\hline
\textbf{Model}        & \textbf{Single Assert} & \textbf{Multiple Asserts} & \textbf{Single + Multi Asserts} \\ \hline
text-davinci & 64.7\%        & 60.7\%           & 62\%                   \\
gemini-pro   & 59.1\%        & 53.8\%           & 55.3\%                 \\
gpt-4.0      & \textbf{67.1\%}        & \textbf{60.9\%}           & \textbf{62.5\% }               \\\hline
\end{tabular}
}
\caption{Accurate match (AM) results across different LLMs.}
\label{tab:discussion}
\end{table}

\han{The \textit{gpt-4.0} achieves higher AM across all fields, likely due to being the latest model from OpenAI. The extended token limits enables it to handle longer test cases and more multiple assertions, which might be challenging for \textit{text-davinci}. A new finding from the non-AM predictions of \textit{gpt-4.0} is it tends to include additional conditions compared to the original ones. For example, where the original assertion is $assert\ A$, \textit{gpt-4.0} might predict $assert\ A\ and\ B$, with $B$ being a valid condition, but not originally specified by the developer.}

\han{The performance of \textit{gemini-pro} is not as good as the two OpenAI models but still outperforms state-of-the-art approaches for single assert prediction (in RQ3). We observed that \textit{gemini-pro} sometimes failed to respond in the requested format, leading to parsing errors and stopped before the feedback interaction phase due to these errors. To enhance the performance of \textit{gemini-pro}, further refinements in prompt design or a more flexible parser for LLM responses may be necessary.}

\han{In summary, the results demonstrate that the design of \tool works well with different models, highlighting its generalisation across different LLMs. Depending on the LLMs, refinements may be necessary for specific models. These refinements might involve minor adjustments in prompt design or response parsing strategies, but do not require major changes to its core structure.}

\subsection{Practical Implications of \tool}
In this section, we discuss two practical usage of \tool: (1) as a wrapper for LLMs, integrated within an IDE or CI process, and (2) to enhance the assert generation capabilities of existing Python unit test generation tools.

\textbf{Usage Scenarios for Developers} \han{\tool can be used as an assistant for developers to help them complete the test cases in their projects. For example, John is a developer for an opensource project. He has just written a new function and is in the process of writing a unit test for the corresponding function. He writes the test inputs but is unsure about the specific assertions needed to thoroughly test the new function. \tool can work as a plugin to his IDE. It automatically analyse the code context and suggest the correct assert statements for John.}

\han{\tool can also be used in continuous integration (CI) environments. When new test code is committed to a repository, \tool can automatically analyse the test code. In cases where the test code may contain bugs or smells (e.g., tests that always pass regardless of the code behavior\footnote{\url{https://github.com/lk-geimfari/mimesis/pull/1385 }}), \tool can compare the generated assert statements with those written by developers. If discrepancies are found, \tool can notify the developers, providing both versions of assert statements and prompting them to decide which one is more appropriate.
Additionally, if the new test code fails, \tool can analyse the code context and suggest alternative assert statements using its feedback interaction mechanism. So if the issue lies within the original assert statements, developers can quickly identify the issue and replace the asserts.}

\textbf{Integration with Automatic Unit Test Generation Tool} Pynguin~\cite{lukasczyk2022pynguin, lukasczyk2023empirical} is a research framework that can automatically generate unit tests for Python and its Github repository has received over 1.1k stars. Due to its design, it requires the target projects to have type annotations. Two of the projects in our dataset overlap with those used in their research, namely Mimesis and Sanic. We employed Pynguin version 0.33.0 with the configuration ``\textit{pynguin --algorithm DYNAMOSA --assertion\_generation MUTATION\_ANALYSIS}'' to generate unit tests for these codebases. We then removed the test cases which were tagged with \textit{pytest.mark.xfail} decorator. These are the cases where Pynguin catch an unexpected error while running the test case. After manual inspect, we found this is due to failed to feed the correct typed parameter of a function. Then, for test cases that failed to produce assertions or the assertions failed, we executed \tool to generate the necessary assertions and subsequently compared the resulting code coverage.

In total, Pynguin generated 98 test cases without the xfail docstring for Sanic and Mimesis. Among them, 25 missed an assert statement. For example, the Pynguin generated cases could have assigned a parameter from the focal method without asserting the parameter.  After execution of \tool, we successfully generated assertions for the untested parameters in all the missing cases and fixed 2 failed test cases where Pynguin generated the wrong assertions. And result in an increased 2\% test coverage for Sanic and Mimesis. These results indicate that \tool can complement Pynguin by rectifying its limitations and enhancing the quality of the automatically generated test cases. Further experiments are necessary to explore how Pynguin can better utilize \tool's capabilities, perhaps by integrating \tool's functionality into its asserts generation process.

Moreover, prior studies such as CODAMOSA~\cite{lemieux2023codamosa} have also explored the potential of LLM models to generate meaningful inputs for the test cases generated from Pynguin, with the intention of mitigating the incidence of test cases decorated with \textit{pytest.mark.xfail} and thus increased the code coverage. By integrating these efforts, there exists prospective future research to enhance the quality of automatically generated unit tests. This could involve the development of a more comprehensive framework that capitalizes on the strengths of LLMs to not only create more meaningful assert statements but also generate the entire test case. It could potentially lead to a improvement in auto-generated unit tests, making them more reliable and beneficial for Python projects.

\subsection{Multiple Asserts and Assert Roulette}

\han{Test smells, similar to code smell, are known for bad design or implementation decisions in the unit test code~\cite{wang2021pynose, van2001refactoring, aljedaani2021test}. Over the years, different test smells in languages such as Java, Javascript, Python were raised by researchers. One of the earliest and most prevalent of these is Assert Roulette~\cite{van2001refactoring}, characterized by test cases that contain multiple assert statements without explanatory messages or documentation~\cite{wang2021pynose}. It can be harmful as developers need more efforts to figure out why the test case failed. It has been one of the most common test smells among languages~\cite{wang2021pynose}. Our randomly sampled data set reveals that 71.8\% of the test cases include more than one assert statement. }

\han{Howerver, recent research by Panichella et al.~\cite{panichella2022test} suggests that the concept of Assertion Roulette may need updating due to advancements in testing frameworks. For example, the unittest framework in Python provides detailed feedback about which asserts fail and the reasons for the failure (e.g., actual vs. expected value). The message is clear enough for the developers to identify the issues. Furthermore,  the choice between using multiple assertions in a single test and spreading them across multiple tests often comes down to a trade-off between assertion roulette and code duplication. Developers often prefer multiple assertions within a single test when these assertions are strongly related, as this approach reduces redundancy. For example, in the Scrapy project, there is a $test\_send\_html$ function\footnote{\url{https://github.com/scrapy/scrapy/blob/10a843ac1d7105d1503ab367e3113da20a2d36fc/tests/test_mail.pyL46}}, two assertions check that both the body and the content type of the sent HTML meet expectations. It is more efficient and clearer to test these aspects together in a single test case rather than duplicating the setup in separate tests.}

\han{Although Assert Roulette is traditionally considered a test smell, it no longer negatively impacts unit tests as it once did, with the evolving testing frameworks and the way developers write unit tests. Therefore, CLAP’s ability to generate multiple assertions within a single test case remains a significant and valuable contribution.}

\section{Threats to Validity}
\label{sec:threat}

\textbf{Internal validity.} A potential threat to internal validity in our study is the extraction of focal methods. We adopted a heuristic approach from previous research~\cite{watson2020learning} to match focal methods with test cases. However, since Python test cases can be structured differently compared to Java, biases may be introduced in the identified focal methods. To address this issue, we modified the script to exclude cases where the focal method was unclear. Additionally, in real-world scenarios, we expect users to be developers who can provide the correct corresponding focal method. Despite this limitation, our approach demonstrates its effectiveness in generating meaningful assert statements, reflecting its robustness.

\textbf{External validity.}
Our study focuses on generating assert statements for Python projects that present potential limitations in external validity, specifically concerning the generalizability of our approach. While the LLMs utilized in our approach have demonstrated versatility across various languages (e.g., CEDAR~\cite{nashidretrieval}), we did not conduct an empirical evaluation on other widely used datasets, such as the Java dataset from ATLAS. This was mainly due to the need for executing real projects, which did not align with the ATLAS database's constraints. Although our method shows the capability to address dynamically-typed languages like Python that possess a greater diversity of assertion types and complex type check compared with Java, the generalization to statically-typed languages remains untested. However, because of the use of the LLM, \tool can potentially be generalized across various programming languages with minor adjustments to the interpreter environment for feedback interaction and the prompt manager design. This indicates a promising scope for \tool's widespread applicability in the assert generation task for a diverse range of programming languages.

\han{While our dataset represents the first dedicated collection for Python assert generation, its 14 projects constructed dataset presents a limitation in generalisability. Despite our efforts to ensure diversity by including projects from various domains (Python libraries, ML/DL repositories, command line tools, and applications) and balancing those using built-in assert keywords with testing libraries, the relatively small number of projects may not capture all Python testing practices. This constraint is due to the large effort required to configure and prepare each project for analysis, and the necessity to avoid data leakage for the LLMs. Future works should aim to expand this dataset to further enhance generalisability of the findings.}

\han{Another potential threat is the adaptability of our prompt design to a variety of LLMs. Given the current popularity of the LLM field, new models are being introduced frequently and old ones may be deprecated (e.g., CODEX). In the experiment and discussion section, we evaluate our approach with four LLMs, and the results demonstrate that all models can generate meaningful assert statements using the current prompt design while a minor modification will be needed for different LLMs. As LLM development continues to evolve, there may be differences in input requirements; however, the core concept of providing natural language prompts to LLMs is expected to remain consistent. Therefore, only limited efforts would be required to adapt our approach to different LLM models.}

\textbf{Construct validity.}
Throughout the study, our aim has been to generate meaningful assert statements; however, the term ``meaningful'' lacks a clear definition. \han{In line with previous studies~\cite{yu2022automated, watson2020learning, nashidretrieval, dinella2022toga, mastropaolo2022using}, we evaluate the results by comparing them with the original asserts from developers. However, it is important to acknowledge that these original asserts may have their own limitations, such as poor readability, ineffectiveness in bug detection, or the presence of test smells. There are other metrics such as the checked coverage~\cite{schuler2011assessing} that can evaluate the quality of the generated test code dynamically rather than the static string-based metrics. However, as Schafer et al.~\cite{schafer2023empirical} pointed out, for dynamically typed language (e.g., JS and Python), it is difficult to precisely implement and calculate the checked coverage.
This could potentially affect the construct validity of our study. We recognise this approach to be an assistant to help developers write assert statements and conduct RQ4 to analyse the differences between the generated asserts and the original ones. Future works could focus on involving a more comprehensive evaluation process and improving the generated assert code quality.}

\section{Related Work}
\label{sec:relatedwork}
\subsection{Assert Generation}

In 2020, Watson et al.~\cite{watson2020learning} first introduced a Deep Learning (DL) method for generating assert statements in Java, referred to as ATLAS. Utilizing a Neural Machine Translation (NMT) based approach, their work aimed to generate meaningful assert statements, which had proven to be a challenge for automatic unit test generation. Moreover, they extracted test cases from Java projects, paired them with their corresponding focal methods, and termed this structure as Test-Assert Pairs (TAPs). They released their dataset for the benefit of future research in the field.

Subsequently, Yu et al.\cite{yu2022automated} extended ATLAS by integrating an Information Retrieval (IR) based approach, while Tufano et al.\cite{tufano2022generating} fine-tuned a state-of-the-art transformer model~\cite{lewis2019bart}, both achieving better performance compared to ATLAS. Similarly, Mastropaolo et al.~\cite{mastropaolo2021studying, mastropaolo2022using} trained T5 model with large Java Dataset and fine-tuned with assert code, also has achieved better performance in assert generation. Most recently, Nashid et al.~\cite{nashidretrieval} combined IR with LLM to generate asserts in Java, achieving state-of-the-art performance. Their tool, CEDAR, which implements the BM25 algorithm to find the most similar test cases from a pre-built dataset and fed into CODEX (an LLM model from OpenAI) as few-shot learning prompts to generate the assert statements. 
In contrast to previous works, which focused on generating single asserts for Java projects, our work addresses the currently most popular language, Python, which has not been previously explored in this context and tends to have more multi-assert test cases. Furthermore, all three previous works relied on pre-built datasets to support their performance, potentially introducing bias and limitations in generating assert statements. Other than Java, 
Zamprogno et al.~\cite{zamprogno2022dynamic} introduced AutoAssert, a system designed to generate assertions for JavaScript. To address the limitations of dynamic approaches in assertion generation, they implemented a human-in-the-loop mechanism by allowing developers to choose the parameters to be tested and select suitable assertions from a predefined set. Like our method, AutoAssert assumes the tested code is correct, but its support is restricted to a limited number of assertion types due to inherent limitations in its design.

\subsection{Large Language Models for Software Engineering}
Recently, various LLMs have been produced by different organizations, such as RoBERTa~\cite{liu2019roberta}, GPT models~\cite{web:gpt}, PaLM~\cite{narang2022pathways}, LLAMA~\cite{touvron2023llama, touvron2023llama2}. In the software engineering domain, researchers and practitioners have conducted numerous experiments to explore the potential contributions of LLMs to the community. 
Copilot~\cite{web:Copilot} assists developers in code writing by reading the code context and comments to generate code suggestions. Codium.AI~\cite{web:codium} utilized LLM to generate unit tests for developers and developed plugins for the IDEs.
CODAMOSA~\cite{lemieux2023codamosa} leveraged OpenAI's CODEX to provide example test cases to help with unit test generation. Xia et al.~\cite{xia2023automated} and Nashid et al.~\cite{nashidretrieval} discussed how the LLM can help with program repair to fix potential bugs. Wang et al.~\cite{wang2022bridging} demonstrated that LLMs exhibit a deeper understanding of source code in tasks like algorithm classification, code clone detection, and code search. And Khan et al.~\cite{khan2022automatic} used LLM to generate meaningful documents for different programming languages. Our project aims to capitalize on the LLM's understanding of source code and incorporate the latest prompt engineering methods to assist developers in generating meaningful assert statements for Python projects.

\section{Conclusion and Future Work}
\label{sec:conclusion}
 In this study, we present \tool, an approach that integrates rounds of prompt communication with persona design, CoT prompting, and feedback interactions to optimize the performance of LLMs in generating meaningful assert statements for Python. Our evaluation demonstrates that \tool outperforms the state-of-the-art approach in producing accurate single assert statements by over 10\%. Adding consideration of multi-asserts generation, which is more common in Python projects, \tool is also capable to generate accurate match assertions at 62\%. 
We further conducted a qualitative study that suggested the generated assertions have the potential to increase the readability of the test case and fix the unseen bugs within the original assertions. Our submitted pull requests based on the generated assertions that positively influenced the test cases were accepted by real open-source projects. In addition, we discuss the generalisation across LLMs, practical applications of \tool, and the importance of multi-asserts generation. \han{In the future, we plan to further refine the method to improve and assess the quality of generated assert statements and minimize test smells. Given the compliant nature of LLMs like the GPT models, we will investigate more robust strategies to ensure these systems not only accept corrections but also apply them in a way that demonstrates a deeper understanding of code exceptions and errors in the prompt engineering aspect. From the practical perspective, we intend to develop this tool into a plugin for Python IDEs to simplify access for developers. Ultimately, we would like to extend the work to automatic unit test generation to better benefit the software engineer community.}


\bibliographystyle{ACM-Reference-Format}
\bibliography{reference}

\end{document}